\documentclass[twocolumn,superscriptaddress,floats,prd,nofootinbib,showpacs]{revtex4}
\usepackage{anyfontsize} 

\usepackage{stmaryrd}
\usepackage{mathrsfs}
\usepackage{float}
\usepackage{amsmath,amssymb,amsfonts}
\usepackage{graphicx}
\usepackage{subfigure}
\usepackage{color} 
\usepackage{fancyhdr}
\usepackage{hyperref} 
\begin{document}

	\title{Quasinormal modes of the Bardeen black hole with a cloud of strings}
	
	\author{Yunlong Liu}
	\affiliation{Department of Physics, South China University of Technology, Guangzhou 510641, China}
	\author{ Xiangdong Zhang\footnote{Corresponding author. scxdzhang@scut.edu.cn}}
	\affiliation{Department of Physics, South China University of Technology, Guangzhou 510641, China}


	\begin{abstract}

	We investigate the quasinormal mode and greybody factor of Bardeen black holes with a string clouds by WKB approximation and verify them by Prony algorithm. We found that the imaginary part of the quasinormal modes spectra is always negative and	the perturbation does not increase with the time, indicating that the system is stable under scalar field perturbation. Moreover, the string parameter $a$ has a dramatically impact on the  frequency and decay rate of the waveforms. In addition, the greybody factor becomes larger when $a$ and $\lambda$ increase while $q$ and $l$ decreases. The parameter $\lambda$ and $l$ have a big effect on the tails. Especially, when $l=0$, a de Sitter phase appears at the tail.
	
	\end{abstract}

	\maketitle

	\section{Introduction}\label{Intro}
	String theory, as we know, is one of the most promising grand unified theories. In string theory, the smallest units of our nature are not point particles, but rather one dimensional extended strings. Due to the inflation in early universe, these fundamental	strings could have been stretched at cosmological sizes \cite{Cosmic_Copeland_2010}. A cloud of strings is the one-dimensional analogous of a cloud of dust. M. Gurses and F. Gursey \cite{Derivation_Gurses_1975} first derived the equations of string motion in general relativity.
	Later,  the solution of Einstein's equation with string cloud is derived by Letelier \cite{Clouds_Letelier_1979} and it is used to establish a star model. Many related papers later considered the string cloud as a fluid in the spacetime background and construct relevant solutions. The physical properties of these black hole solutions have been investigated \cite{Cloud_Ghosh_2014, Cloud_Ghosh_2014a, Quasinormal_Graca_2017, Black_Toledo_2018, Reissner_Toledo_2019, Clouds_Singh_2020}. Take Schwarzschild black hole with a cloud of string for instance, the event horizon radius receives a correction as $r_h=2M/(1-a)$ with $a$ being string clouds parameter \cite{Clouds_Letelier_1979}. This modification factor $1/(1-a)$ may have some potential astrophysical applications \cite{Radiation_Glass_1998,Traversable_Richarte_2008}.

	On the other hand, in general, there is always a singularity in the black hole solution, which is enveloped by the event horizon. In contrast, Bardeen proposed a black hole solution without a curvature singularity \cite{Nonsingular_Bardeen_1968}. Beato and Garcia \cite{Bardeen_Ayon-Beato_2000} proposed a magnetic solution of the Einstein equations coupled with nonlinear electrodynamics. Then, several paper have studied this type of black hole solution \cite{Quantum_Sharif_2010, Quasinormal_Fernando_2012, Bardeen_Rodrigues_2018, Bardeen_Rodrigues_2022, BardeenKiselev_Rodrigues_2022}.
	Bardeen black hole solution with a cloud of strings was obtained very recently in \cite{Bardeen_Rodrigues_2022}. Though this black hole solution has the same event horizon characteristic as the regular Bardeen solution. However, the strings parameter $a$ make the solution singular at the origin. Due to the such significant changes in the properties of black holes, investigating the various intrinsic characters of this black hole becomes an interesting topic.

	It is well known that one powerful way to extract the black hole characterization is to perturb it and then see its response. When considering a perturbation that can be ignored in the spacetime background and choosing an appropriate gauge, the evolution of this perturbation can be described by a series of simple wave equations \cite{Stability_Regge_1957,Effective_Zerilli_1970}.
	Through these perturbation wave equations, we can see that the evolution of perturbation mainly consists of three stages: initial outburst, quasinormal ringing, and finally the power law (asymptotically flat spacetime) or exponential (asymptotic de Sitter spacetime) tail \cite{Blackhole_Maggiore_2018}.
	The quasinormal ringing stage, where quasinormal modes (QNMs) are mainly determined by the parameters of the black hole and are independent of the initial perturbation \cite{Spectral_Leaver_1986, Quasinormal_Berti_2009}, is an important component of current gravitational wave detection \cite{LIGO_Abbott_2016}.
	For the third stage, it is mainly caused by the scattering of perturbations at infinity \cite{Wave_Ching_1995}. This stage is of great significance for studying the stability of black holes. Currently, a large number of papers have studied QNMs and power law tails in different spacetime backgrounds by various methods \cite{Quasinormal_Konoplya_2011,Zhang12,Quasinormal_Graca_2017,Conformal_Konoplya_2021,Quasinormal_Xiong_2022,Instability_Yang_2022,Quasinormal_Fu_2023}. For example, a recent paper \cite{Conformal_Konoplya_2021} studied the QNMs of a Schwarzschild-like black holes with cosmological constant in conformal Weyl gravity and found that the evolution of scalar field is divided into three stages: Schwarzschild ringing stage, effective dark matter ringing stage, and an exponential tail of de Sitter stage. Moreover, \cite{Quasinormal_Fernando_2012} analyzed the QNMs of a Bardeen black hole caused by scalar perturbations and compared them with results in the Reissner-Nordstrom black hole.

	Given the above motivations, as the first step toward understanding the properties of this Bardeen black hole
	with a cloud of strings (SBBH), we consider a probe massless scalar field over this background and study the
	properties of its QNM spectra in this paper.

	The structure of this paper is outlined as follows:
	In Sec. \ref{SandW},  the metric of SBBH and the scalar perturbation in this background are introduced, and the corresponding effective potential is given.
	In Sec. \ref{method}, the methods used in this paper to analyze QNMs and greybody factors are introduced, including the finite element method(FEM), WKB approximation, Prony algorithm.
	The effects of various parameters on QNMs and greybody factors are calculated for SBBH in Sec. \ref{result} and then the accuracy of the calculations is verified by the Prony method in Sec. \ref{PronyTest}.
	Finally, some conclusions and corresponding discussions are given in Sec. \ref{conclusion}.

	\section{Spacetime and wave equation}\label{SandW}

	\subsection{Spacetime}\label{spacetime}
	As for SBBH, the action can be described as the general relativity minimally coupled to the nonlinear electrodynamics (NED) and string clouds as follows \cite{Bardeen_Rodrigues_2022}:
	\begin{eqnarray}
		S=\int d^{4} x \sqrt{-g}[R+2 \lambda+\mathcal{L}(F)]+S_{C S}.
	\end{eqnarray}
	Here, $R$ is the Riemann scalar, $\lambda$ is the cosmological constant, $S_{CS}$ is the Nambu-Goto action \cite{Clouds_Letelier_1979} used to describe string-like objects,
	as
	\begin{eqnarray}
		S_{S C}=\int \mathcal{M}\left(-\frac{1}{2} \Sigma^{\mu \nu} \Sigma_{\mu \nu}\right) d \Lambda^{0} d \Lambda^{1},
	\end{eqnarray}
	where $\Lambda^{0}$ is a timelike parameter while $\Lambda^{1}$ is a spacelike one, and the string  cloud parameter $\mathcal{M}$ is a dimensionless constant.  $\Sigma^{\mu \nu}$  is given by
	\begin{eqnarray}
		\Sigma^{\mu \nu}=\epsilon^{A B} \frac{\partial x^{\mu}}{\partial \Lambda^{A}} \frac{\partial x^{\nu}}{\partial \Lambda^{B}} .
	\end{eqnarray}
	The Levi-Civita symbol $\epsilon^{A B}$ meets $\epsilon^{01}=-\epsilon^{10}=1$.

	Analogously, $\mathcal{L}(F)$ is the Lagrangian corresponding to Bardeen's solution \cite{Quasinormal_Fernando_2012}:
	\begin{eqnarray}
		\mathcal{L}(F)=\frac{3}{8 \pi s q^{2}}\left(\frac{\sqrt{2 q^{2} F}}{2+\sqrt{2 q^{2} F}}\right)^{5 / 2},
	\end{eqnarray}
	where the scalar $F=F^{\mu\nu}F_{\mu\nu}/4$, $q$ is the magnetic charge, $M$ is the mass of the magnetic monopole, and $s=|q|/(2M)$. The static spherically symmetric solution to this theory was given as follows \cite{Bardeen_Rodrigues_2022}:
	\begin{eqnarray}
		ds^2=f(r) dt^2-\frac{1}{f(r)}dr^2-r^2 d\theta^2-r^2\sin^2{\theta}d\phi^2,\label{SBBH}
	\end{eqnarray}
	where
	\begin{eqnarray}
		f(r)=1-a-\frac{2 M_{1}}{r}-\frac{2 M r^{2}}{\left(q^{2}+r^{2}\right)^{3 / 2}}-\frac{\lambda r^{2}}{3}.
	\end{eqnarray}
	
	Here, $a$ is an integration constant related to the string, with a constraint range of $0<a<1$. $M_1$ is an integration constant generated during the solution of the differential equation, and $M_1$ is usually set to zero.
	When $a=0$ and $\lambda=0$, this spherically symmetric spacetime can be returned to the Bardeen black hole solution. Though this solution \eqref{SBBH} has the similar event horizon characteristic as the Bardeen solution. However, different with the regular Bardeen solution, the existence of the parameter of the strings $a$ make the solution singular at the origin \cite{Bardeen_Rodrigues_2022}. Moreover, in this paper, we restrict ourself for positive $\lambda$ (de Sitter case).
	
	\subsection{Wave equation}\label{wavefun}
	
	The motions of massless scalar field $\psi$ in SBBH can be described by the Klein-Gordon (KG) equation:
	\begin{eqnarray}
		\frac{1}{\sqrt{-g}} \partial_{\mu}\left(\sqrt{-g} g^{\mu \nu} \partial_{\nu} \psi\right)=0.\label{KGequation}
	\end{eqnarray}
	
	Through the separation of variables by spherical harmonic functions, $\psi=\Psi(t,r)Y_l(\theta,\phi)/r$, the Eq. \eqref{KGequation} can be reduced to:
	\begin{eqnarray} \label{timedomain}
		-\frac{\partial^{2} \Psi}{\partial t^{2}}+\frac{\partial^{2} \Psi}{\partial r_{*}^{2}}-V(r) \Psi=0.
	\end{eqnarray}
	
	Further separating the time variable, assuming $\Psi=e^{-i \omega t} \Phi$, the Eq. \eqref{timedomain} further simplifies to
	\begin{eqnarray} \label{fredomain}
		\frac{\partial^{2} \Phi}{\partial r_{*}^{2}}+\left(\omega^{2}-V(r)\right) \Phi=0.
	\end{eqnarray} Here the effective potential reads
	\begin{eqnarray}
		V(r) = f(r)\left(\frac{l(l+1)}{r^2} + \frac{1}{r}  \frac{df(r)}{dr}\right),
	\end{eqnarray}where $l$ is the angular quantum number and $r_*$ is the tortoise coordinate defined as $dr_{*}=dr/f(r)$. In the Bardeen solution, due to the complexity of the function $f(r)$, we usually cannot obtain an explicit solution $r_*(r)$. Therefore, in our calculation, we use the method of numerical integration and interpolation to obtain a solution, see Appendix \ref{Append1}.
	
	
	\section{QNMs analysis: through WKB approximation and Finite Element Method}\label{method}
	Calculating QNMs is essentially obtaining the intrinsic frequencies of Eq. \eqref{fredomain}. In order to solve this equation, some boundary conditions are required. For cases with cosmological constant, the constraint conditions near the event horizon and cosmological horizon require that the waves propagate towards these horizons, respectively. While for cases without cosmological constant, the boundary condition is to require that the wave solutions propagate outward at infinite spatial distance.
	
	However, even with these constraints, the wave equation for black hole perturbations usually can not analytically solvable. Therefore, many numerical methods have been developed to calculate QNMs for different systems \cite{Quasinormal_Konoplya_2011}.
	
	In this section, we first introduce the WKB approximation method for calculating QNMs. Then, we describe Finite Element Method (FEM) for solving the wave equation in SBBH with a given initial perturbation, and obtain the evolution in  time domain. In addition, we describe the Prony method for extracting the QNMs with $n=0$ from the scalar evolution data. Finally, we briefly introduce the greybody factor in the WKB approximation.

	\subsection{WKB Approximation}
	The method of using WKB approximation to solve QNMs was first proposed by Schutz and Will \cite{Black_Schutz_1985}. This method is suitable for calculating effective potentials with potential barriers and constant values  near the boundaries.
	Later, Iyer and Will \cite{Blackhole_Iyer_1987} obtained the 3rd-order WKB approximation, which was further improved by Konoplya \cite{Quasinormal_Konoplya_2003} to the 6th-order WKB approximation, and this method was used to calculate QNMs of the D-dimensional Schwarzschild black hole.
	Recently, Matyjasek and Opala \cite{Quasinormal_Matyjasek_2017} combined Pade approximation to improve the accuracy up to the 13th order.
	For the WKB approximation method, it can be described uniformly \cite{Higher_Konoplya_2019} as follows:
	\begin{eqnarray} \label{WKBFun}
		\omega^{2}
		&=&
		V_{0}+A_{2}\left(\mathcal{K}^{2}\right)+A_{4}\left(\mathcal{K}^{2}\right)+A_{6}\left(\mathcal{K}^{2}\right)+\ldots \notag  \\
		&& -\mathrm{i} \mathcal{K} \sqrt{-2 V_{2}}\left(1+A_{3}\left(\mathcal{K}^{2}\right)+A_{5}\left(\mathcal{K}^{2}\right)+\ldots\right),
	\end{eqnarray}
	where $\mathcal{K}=\pm (n+1/2)$, $A_{i}\left(\mathcal{K}^{2}\right),i=2,3...$ are the modifications for the $i$th order. $V_{i},i=0, 2,3...$ are the values of $V(x)$ and higher order derivatives at the maximum value.
	Note that it is necessary to estimate errors by comparing differences between different orders. The error estimation $\Delta_{k}$ for the $k$-th order WKB approximation can be expressed as:
	\begin{eqnarray}
		\Delta_{k}=\frac{\left|\omega_{k+1}-\omega_{k-1}\right|}{2}	,
	\end{eqnarray}
	where $\omega_{k}$ represents the QNMs obtained from the k-th order WKB approximation.
	It should be noted that the higher order of WKB approximation does not necessarily lead to higher accuracy \cite{Quasinormal_Hatsuda_2020}. Therefore, Pade approximation is usually used to improve the accuracy of high-order WKB approximations.
	In this paper, we will use the 6th-order WKB approximation with Pade approximation to perform calculation analysis.
	
	\subsection{Finite Element Method}
	Given the initial perturbation, we can obtain the dynamic evolution of the initial perturbation through the wave equation. To obtain the dynamic evolution, we use the FEM. It replaces continuous differentials with a series of discrete differences. The differential equation \eqref{timedomain} can be replaced by:
	\begin{eqnarray} \label{FDMFun}
		\Psi_{j}^{i+1}=
		&&-\Psi_{j}^{i-1}
		+\left(2-2 \frac{\Delta t^{2}}{\Delta r_{*}^{2}}-\Delta t^{2} V_{j}\right) \Psi_{j}^{i}   \notag \\
		&&+\frac{\Delta t^{2}}{\Delta r_{*}^{2}}\left(\Psi_{j-1}^{i}+\Psi_{j+1}^{i}\right) ,
	\end{eqnarray}
	where $t_{i}=t_{0}+i \Delta t$, $r_{* j}=r_{* 0}+j \Delta r_{*}$, and $\Psi_{j}^{i}=\Psi\left(t=t_{i}, r_{*}=r_{* j}\right)$, $V_{j}=V\left(r_{*}=r_{* j}\right)$.
	The initial conditions are chosen as:
	\begin{eqnarray}
		&&\Psi\left(r_{*}, t_{0}\right)=C_{A} \exp \left(-C_{a}\left(r_{*}-C_{b}\right)^{2}\right), \\
		&&\left.\frac{\partial}{\partial t} \Psi\left(r_{*}, t\right)\right|_{t=t_{0}}=0 .
	\end{eqnarray}
	To satisfy the von Neumann stability condition \cite{Scalar_Lin_2016}, we choose $\Delta t/\Delta r_{*}=2/3$, and ensure that $\Delta t$ is small enough.
	
	\subsection{Prony Method}\label{PronyM}
	The Prony method is an analysis technique for extracting signal phase, frequency, amplitude, and damping coefficients from the time domain. In this method, frequency and damping coefficients correspond to the real and imaginary parts of the QNMs. We assume that the signal is composed of a series of damped sinusoidal signals, which can be simply described as \cite{Mining_Berti_2007}:
	\begin{eqnarray}\label{Pronyfit}
		\Psi(t) \approx \Sigma_{j=1}^{p} C_{j} e^{-i \omega_{j} t}.
	\end{eqnarray}
	By combining appropriate data and conducting numerical analysis, we can obtain the QNM frequencies $\omega_{j}$ that we need.
	Generally speaking, the fundamental mode signal, i.e., overtone index $n=0$, has the longest lifetime in QNM signals. Other signals ($n>0$) will disappear due to rapid decay. For example, from Tables \ref{aQNM} to \ref{lQNM}, we can see that the decay rate at $n=1$ is generally faster than that at $n=0$. Therefore, we mainly use the Prony analysis method to extract the fundamental frequency of the QNMs. Note that when using the Prony method to extract fundamental frequency information, we usually choose a time period after the QNM signals arrive and before the onset of the tail.
	
	\subsection{Greybody Factor}\label{GrayBody}
	In this Part, we introduce the scheme of using the WKB method to analyze the greybody factor which can be used to further describe the intrinsic characteristics of the effective potential of the background spacetime.
	
	For the wave Eq. \eqref{fredomain}, we consider the scattering boundary conditions:
	\begin{eqnarray}
		\Phi&=&T e^{-i \omega r_{*}}, \quad r_{*} \rightarrow-\infty, \\
		\Phi&=&e^{-i \omega r_{*}}+ R e^{i \omega r_{*}}, \quad r_{*} \rightarrow \infty,
	\end{eqnarray}
	where $T$ is the transmission coefficient and $R$ is the reflection coefficient. In particular, when the effective potential is real, in Eq. \eqref{WKBFun}, $\mathcal{K}$ is a purely imaginary constant, and its relationship with the reflection coefficient and transmission coefficient are as follows \cite{Blackhole_Iyer_1987}:
	\begin{eqnarray}
		|R|^{2}  &=&\frac{1}{1+\mathrm{e}^{-2 \pi \mathrm{i} \mathcal{K}}}, \quad 0<|R|^{2}<1 , \\
		|T|^{2}  &=&\frac{1}{1+\mathrm{e}^{2 \pi \mathrm{i} \mathcal{K}}}=1-|R|^{2} .
	\end{eqnarray}
	It is worth noting that the eikonal formula gives an approximate solution for $\mathcal{K}$:
	
	\begin{eqnarray}
		\mathcal{K}=- \frac{ V_{0}- \omega^2}{\mathrm{i} \sqrt{- 2 V_{2}}},
	\end{eqnarray} and other terms in Eq. \eqref{WKBFun} can be considered as higher-order corrections.
	
	\section{Calculation of Time-Domain Solution and QNMs}\label{result}
	Before analyzing the influence of various parameters on QNMs, we first analyze the evolution characteristics of initial scalar perturbations in $(t,r_*)$ spacetime diagram, as shown in the following figure:
	\begin{figure}[!htb]
		\includegraphics [width=0.44\textwidth]{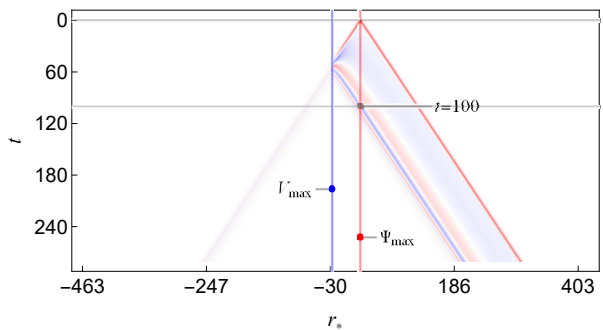}
		\caption{Evolution of perturbed scalar field $\Psi$ over time.}
		\label{PsiTimeAll}
	\end{figure}
	
	In this spacetime diagram of perturbed scalar field $\Psi$, the parameters we have chosen are:
	$$
	\left\{ M \rightarrow \frac{1}{2}, a \rightarrow \frac{1}{5}, q \rightarrow \frac{1}{5}, \lambda \rightarrow \frac{1}{500}\right\}\{l \rightarrow 2\}.
	$$
	The initial perturbations are selected as:
	$$C_A=10; \quad C_a=1/8;\quad C_b=r_*(V_{max})+50.$$
	We use this initial perturbation for all subsequent calculations.
	
	From Fig. \ref{PsiTimeAll}, we can see that on the left side of the effective potential maximum $V_{max}$, the evolution of the scalar field has only two stages: one is the QNM stage transmitted from the barrier, and the other is the tail stage after some time. On the right side of the effective potential maximum $V_{max}$, as mentioned in the introduction, the evolution of the scalar field is divided into three stages (i.e., initial burst, quasinormal ringing, and tail). For example, at $\Psi_{max}$, because the wave speed of the perturbation is $1$ and $r_*(\Psi_{max})-r_*(V_{max})=50$, we can infer that the arrival time of quasinormal ringing is $t=100$. Hence, At $\Psi_{max}$, the time intervals corresponding to the three stages of the scalar field evolution are shown in the figure: initial burst $(0<t<100)$, quasinormal ringing and tail $(t>100)$.
	
	\subsection{Influence of parameter \texorpdfstring{$a$}{a}}
	In this section, we analyze the influence of parameter $a$ on QNMs. First, we fix all parameters except for $a$, which are set to
	$$\left\{ M \rightarrow \frac{1}{2}, q \rightarrow \frac{1}{5}, \lambda \rightarrow \frac{2}{1000}\right\}\{l \rightarrow 2\},$$
	while $a$ takes four different values, $a= 0, 0.2, 0.4, 0.6$.
	\begin{figure}[!htb]
		\includegraphics [width=0.45\textwidth]{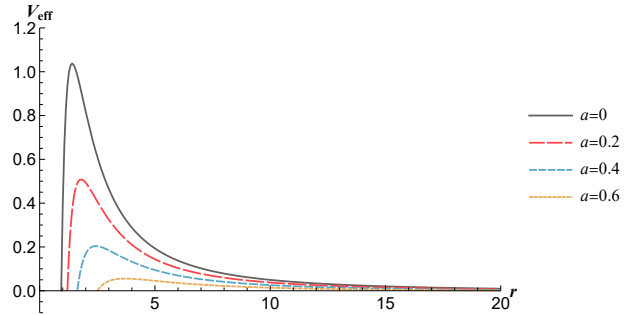}
		\caption{The effective potential $V$ for different values of $a$.}
		\label{aVeff}
	\end{figure}

	\begin{figure}[!htb]
		\includegraphics [width=0.45\textwidth]{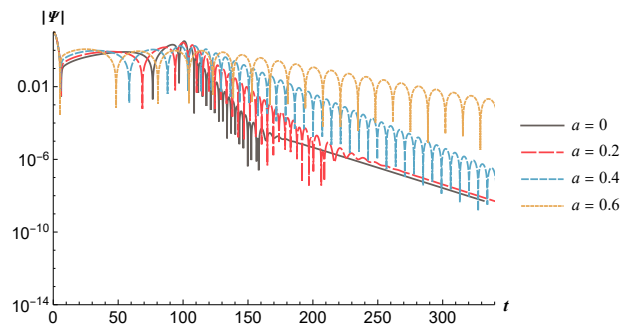}
		\caption{ The evolution diagram of scalar field perturbation $\Psi$ for different values of $a$.}
		\label{aPsi}
	\end{figure}
	Fig. \ref{aVeff} shows the effective potential $V$ outside the event horizon for  different values of $a$. We can see that the peak of the effective potential decreases and the curvature at the peak becomes smaller as $a$ increases.
	
	Generally speaking, the larger the peak of the effective potential, the larger the peak of the scalar field scattered by the effective potential. In the previous analysis, it was estimated that the reflected scalar field perturbation reaches its peak at $t=100$. Fig. \ref{aPsi} shows that at $t=100$, as $a$ increases and the peak of the effective potential gradually decreases, the peak of $|\Psi|$ indeed decreases correspondingly.
	
	Similarly, the smaller the curvature at the peak of the effective potential, the lower the fundamental frequency of the QNMs of the scalar field scattered by the effective potential. Fig. \ref{aPsi} indeed shows that as $a$ increases and the curvature at the peak of the effective potential decreases, the oscillation frequency of $|\Psi|$ at $t>100$ also decreases significantly.

Fig. \ref{aPsi} shows that the string parameter $a$ has a dramatically impact on the  frequency and decay rate of the waveforms. The decay rate becomes slower as $a$ increases.
	
	To further verify our findings,  the QNMs obtained by the WKB approximation is given in Table \ref{aQNM}.
	We can see that as $a$ increases, the real part of the fundamental frequency of the QNMs decreases, while the absolute value of the imaginary part also decreases.
	
	\begin{table}[!htb]
	\caption{QNMs for different values of $a$ in the SBBH.}
	\begin{ruledtabular}
		\begin{tabular}{llll}
		$a$ &$n$& Quasinormal frequency		  & error estimation		  \\
		\hline
		0	& 0 & 0.9935109586-0.1888357704 i & $1.273609 \times 10^{-6}$ \\
		 	& 1 & 0.9589280861-0.5756241876 i & $2.679108 \times 10^{-5}$ \\
		0.2 & 0 & 0.6980498152-0.1216230302 i & $3.549236 \times 10^{-7}$ \\
			& 1	& 0.6770331167-0.3698417504 i & $9.113942 \times 10^{-6}$ \\
		0.4	& 0 & 0.4446334188-0.0683830206 i & $5.856024 \times 10^{-8}$ \\
			& 1 & 0.4340341896-0.2072901971 i & $6.711971 \times 10^{-7}$ \\
		0.6	& 0 & 0.2331775703-0.0297377604 i & $5.458729 \times 10^{-9}$ \\
			& 1	& 0.2294816421-0.0897878689 i & $4.080292 \times 10^{-8}$ \\
			\end{tabular}
		\end{ruledtabular}
	\label{aQNM}
	\end{table}
	This means that as $a$ increases, the oscillation frequency in the quasinormal ringing will decrease and the decay rate will slow down. This is consistent with the results shown in Fig. \ref{aPsi}.
	\begin{figure}[!htb]
		\includegraphics [width=0.45\textwidth]{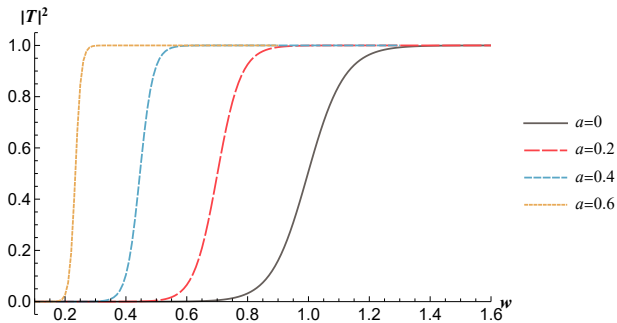}
		\caption{graybody factor $|T|^2$ for  varying $a$.}
		\label{aGraybody}
	\end{figure}
	Finally, we present the corresponding graybody factors in Fig. \ref{aGraybody}.
	We can see that as $a$ increases, the transmittance of the black hole horizon gradually increases, which is consistent with the decrease of the peak of the effective potential.

	\subsection{Influence of the magnetic charge \texorpdfstring{$q$}{q}}
	In this part, we analyze the influence of $q$ on QNMs. Similar to the analysis of the parameter $a$, we first fix all other parameters except for $q$ as follows:
	$$\left\{ M \rightarrow \frac{1}{2}, a \rightarrow \frac{1}{5}, \lambda \rightarrow \frac{2}{1000}\right\}\{l \rightarrow 2\}.$$
	\begin{figure}[!htb]
		\includegraphics [width=0.45\textwidth]{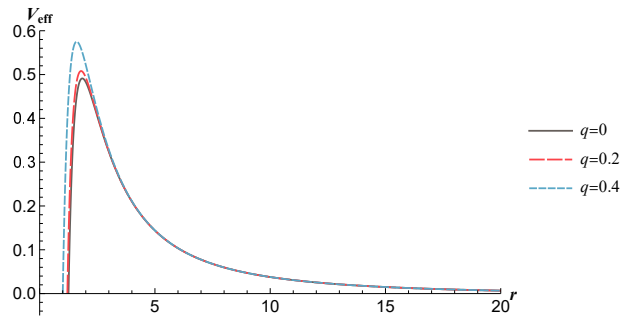}
		\caption{The effective potential $V$ for different values of $q$.}
		\label{qVeff}
	\end{figure}
	
	\begin{figure}[!htb]
		\includegraphics [width=0.45\textwidth]{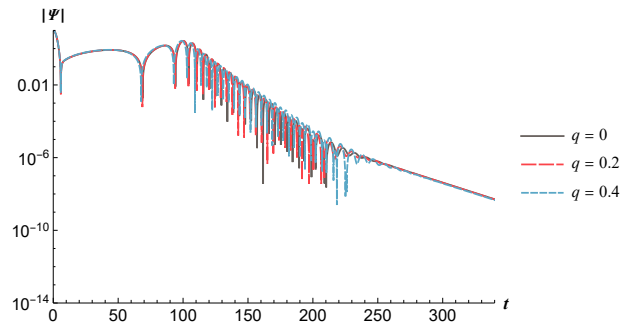}
		\caption{ The evolution diagram of scalar field perturbation $\Psi$ for different values of $q$.}
		\label{qPsi}
	\end{figure}
	Fig. \ref{qVeff} shows the variation of effective potential $V$ for $q = 0, 0.2, 0.4$ outside the event horizon. Similarly, between the event horizon and cosmological horizon, the effective potential is always greater than zero. The figure shows that the maximum value of the effective potential $V$ increases with the increasing of the magnetic charge $q$. However, this change is very small. It can be seen that the effect of $q$ is only significant near the event horizon. When $r$ is large enough, $q$ has a very weak effect on the the effective potential $V$. Therefore, near the cosmological horizon, the effective potentials are almost identical in Fig. \ref{qVeff}.
	
	The insignificant change, especially near the cosmological horizon, in the effective potential leads to an unremarkable change in the evolution of scalar field. As shown in Fig. \ref{qPsi}, it can be seen that the evolution of the scalar field almost overlaps.
	
	\begin{table}[!htb]
		\caption{QNMs for different values of $q$ in the SBBH.}
		\begin{ruledtabular}
			\begin{tabular}{llll}
		$q$ & $n$ & Quasinormal frequency & error estimation \\
		\hline
		0	& 0	& 0.68542516-0.12323891 i & $3.089469 \times 10^{-7}$ \\
		 	& 1	& 0.66268336-0.37512830 i & $6.171675 \times 10^{-7}$ \\
		0.2	& 0 & 0.69804982-0.12162303 i & $3.549236 \times 10^{-7}$ \\
			& 1	& 0.67703312-0.36984175 i & $9.113941 \times 10^{-6}$ \\
		0.4	& 0 & 0.74594502-0.11285594 i & $5.491912 \times 10^{-7}$ \\
			& 1	& 0.72949588-0.34154418 i & $8.846580 \times 10^{-6}$ \\
			\end{tabular}
		\end{ruledtabular}
	\label{qQNM}
	\end{table}
	Similarly, we can analyze the QNMs in more detail, using the WKB approximation.
	Table \ref{qQNM} shows the variation of QNMs with $q$. It can be seen that the larger $q$ cause the higher value of the fundamental frequency $Re(\omega)$ of QNMs, while the opposite trend on the value of $|Im(\omega)|$. This means that in the Quasinormal ringing, as $q$ increases, the oscillation frequency will increase, and the decay rate will slow down correspondingly in Fig. \ref{qPsi}.
	
	\begin{figure}[!htb]
		\includegraphics [width=0.45\textwidth]{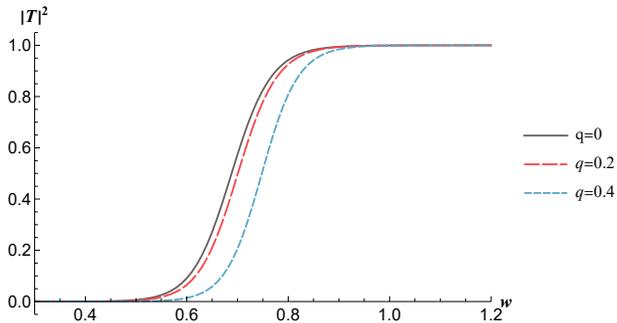}
		\caption{graybody factor $|T|^2$ for  varying $q$.}
		\label{qGraybody}
	\end{figure}
	Finally, in Fig. \ref{qGraybody}, we give the corresponding graybody factor. It shows that when the frequency $\omega$ is fixed, as parameter $q$ increases, the graybody factor $|T|^2$ decreases, synchronously.

	\subsection{Influence of cosmological constant \texorpdfstring{$\lambda$}{}}
	The influence of cosmological constant $\lambda$ on QNMs was analyzed in this section. Similarly, we first determine the values of parameters ($M$, $a$, $q$, $l$) as follows:
	$$\left\{ M \rightarrow \frac{1}{2}, a \rightarrow \frac{1}{5}, q \rightarrow \frac{1}{5} \right\}\{l \rightarrow 2\}.$$
	
	\begin{figure}[!htb]
		\includegraphics [width=0.45\textwidth]{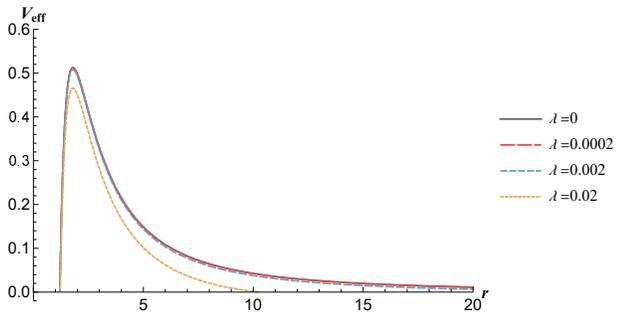}
		\caption{The effective potential $V$ for different values of $\lambda$.}
		\label{lambdaVeff}
	\end{figure}
	
	\begin{figure}[!htb]
		\includegraphics [width=0.45\textwidth]{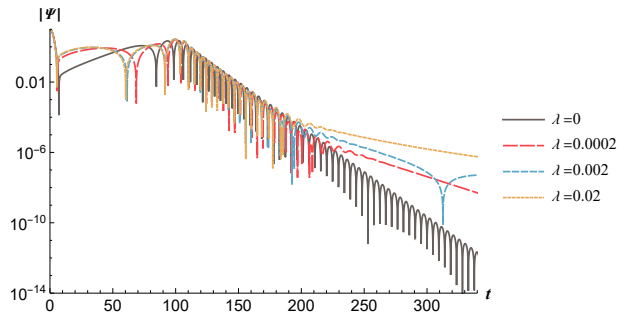}
		\caption{The evolution diagram of scalar field perturbation $\Psi$ for different values of $\lambda$.}
		\label{lambdaPsi}
	\end{figure}

	In Fig. \ref{lambdaVeff}, we show how the effective potential $V$ outside the event horizon changes as the parameter $\lambda$ varies. We can see that as the cosmological constant $\lambda$ increases, the peak value of the effective potential decreases correspondingly.
	
	In addition, unlike the $q$ which primarily affects the behavior of $V$ near the event horizon, $\lambda$ mainly affects $V$ near the cosmological horizon (while at infinity for $\lambda=0$).
	Meanwhile, the tail part of the $\Psi$ is mainly determined by the behavior of the effective potential near the cosmological horizon \cite{Wave_Ching_1995}.
	
	In Fig. \ref{lambdaPsi}, we can see that with the increase of the parameter $\lambda$, the tail of $|\Psi|$ does undergo significant changes. Compared to the almost unchanged tail caused by parameter $q$, this proves the point in \cite{Wave_Ching_1995}.
	
	Table \ref{lambdaQNM} shows the QNMs obtained by the WKB approximation for the different value of $\lambda$.
	\begin{table}[!htb]
		\caption{QNMs for different values of $\lambda$ in the SBBH.}
		\begin{ruledtabular}
			\begin{tabular}{llll}
		$\lambda$  & $n$ & Quasinormal frequency & error estimation \\
		\hline
		0 		& 0 & 0.70122581-0.12208316 i & $4.601346 \times 10^{-7}$ \\
		  		& 1 & 0.67987976-0.37136248 i & $9.182041 \times 10^{-6}$ \\
		0.0002 	& 0 & 0.70090873-0.12203731 i & $4.464850 \times 10^{-7}$ \\
				& 1 & 0.67959576-0.37121074 i & $9.178675 \times 10^{-6}$ \\
		0.002 	& 0 & 0.69804982-0.12162303 i & $3.549236 \times 10^{-7}$ \\
				& 1 & 0.67703312-0.36984175 i & $9.113941 \times 10^{-6}$ \\
		0.02    & 0 & 0.66892071-0.11731892 i & $3.007880 \times 10^{-7}$ \\
		 		& 1 & 0.65073136-0.35581359 i & $7.112285 \times 10^{-6}$ \\
			\end{tabular}
		\end{ruledtabular}
	\label{lambdaQNM}
	\end{table}
	In Table \ref{lambdaQNM}, the absolute value of the real and imaginary part of the fundamental frequency of the QNMs decreases as $\lambda$ grows.

	\begin{figure}[!htb]
		\includegraphics [width=0.45\textwidth]{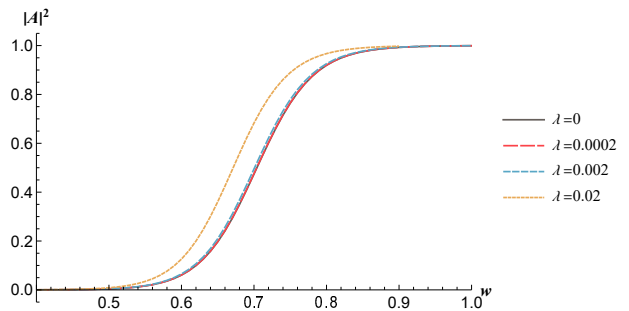}
		\caption{graybody factor $|T|^2$ for  varying $\lambda$.}
		\label{lambdeGraybody}
	\end{figure}
	Finally, in Fig. \ref{lambdeGraybody}, the graybody factor $|T|^2$ decreases as the cosmological parameter increases for a fixed frequency $\omega$. This corresponds to the change of  the peak of $V$ with $\lambda$.
	
	\subsection{Influence of  \texorpdfstring{$l$}{l}}
	In this section, we analyzed the influence of parameter $l$ on the QNMs. We first fixed the parameters that were not under consideration as follows:
	$$\left\{ M \rightarrow \frac{1}{2} , a \rightarrow \frac{1}{5}, q \rightarrow \frac{1}{5}, \lambda \rightarrow \frac{2}{1000}\right\}.$$
	
	\begin{figure}[!htb]
		\includegraphics [width=0.45\textwidth]{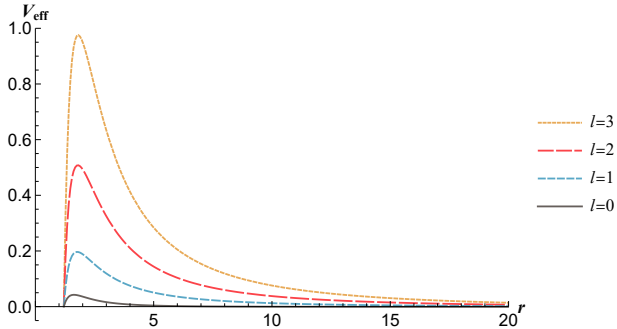}
		\caption{The effective potential $V$ for $l=0, 1, 2, 3$.}
		\label{lVeff}
	\end{figure}

	\begin{figure}[!htb]
		\includegraphics [width=0.45\textwidth]{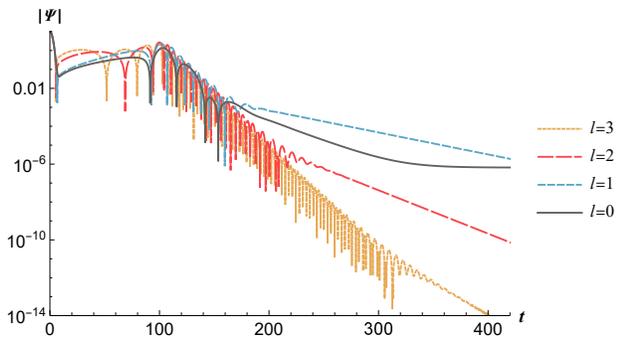}
		\caption{The evolution diagram of scalar field perturbation $\Psi$ for $l=0, 1, 2, 3$.}
		\label{lPsi}
	\end{figure}
	In Fig. \ref{lVeff}, we demonstrate the variation of the effective potential $V$ outside the event horizon with respect to $l$. As $l$ increases, the peak value of the effective potential rises accordingly, while the curvature at the peak also increases. This behavior is similar to that of $a$. Correspondingly, during the quasinormal ringing phase $100<t<t_t$ (where $t_t$ denotes the start of the tail phase) shown in Fig. \ref{lPsi}, we observe that the oscillation frequency of $\Psi$ increases with $l$, and that the amplitude of $\Psi$ slightly grows around $t=100$ with increasing $l$.

	Interestingly, in Figure \ref{lPsi}, for the black solid line when $l=0$, there appears a nearly flat tail around $t\approx 320$, which corresponds to the de Sitter phase as described in reference \cite{Conformal_Konoplya_2021,Radiative_Brady_1999,Late_Ismail_2021}.
	
	It is noteworthy that when $l>0$, the effective potential remains greater than zero between the event horizon and the cosmological horizon. However, when $l=0$, the effective potential no longer represents a mere potential barrier but rather a potential well. For example, $V_{eff}(20)=-0.000584$. It implies that the physical system may harbor bound states.
	
 	Next, we furnish detailed results of QNMs obtained by WKB approximation in Table \ref{lQNM} for $l=0, 1, 2, 3$.
	\begin{table}[!htb]
		\caption{QNMs for different values of $l$ in the SBBH.}
		\begin{ruledtabular}
			\begin{tabular}{llll}
		$l$ & $n$ & Quasinormal frequence & error estimation \\
		\hline
		0 & 0 & 0.14440104-0.13157036 i & $6.7648323 \times 10^{-5}$ \\
		  & 1 & 0.11650529-0.44025206 i & $3.5981065 \times 10^{-4}$ \\
		1 & 0 & 0.41910739-0.12260593 i & $3.7594813 \times 10^{-6}$ \\
		  & 1 & 0.38788734-0.38077137 i & $4.5168872 \times 10^{-5}$ \\
		2 & 0 & 0.69804982-0.12162303 i & $3.5492360 \times 10^{-7}$\\
		  & 1 & 0.67703312-0.36984175 i & $9.1139415 \times 10^{-6}$\\
		3 & 0 & 0.97715823-0.12134890 i & $5.0536923 \times 10^{-8}$\\
		  & 1 & 0.96168430-0.36662325 i & $1.3181977 \times 10^{-6}$ \\
			\end{tabular}
		\end{ruledtabular}
	\label{lQNM}
	\end{table}
	From Table \ref{lQNM}, we observe that as $l$ increases, the real part of the fundamental frequency of QNMs gradually increases, while the absolute value of the imaginary part decreases. Thus, during the quasinormal ringing phase, the oscillation frequency increases and the decay rate slows down with the rise of $l$ as shown in Fig. \ref{lPsi}.
	
	\begin{figure}[!htb]
		\includegraphics [width=0.45\textwidth]{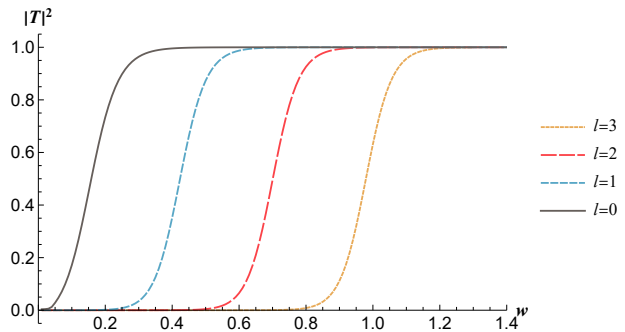}
		\caption{graybody factor $|T|^2$ for  varying  $l$.}
		\label{lGraybody}
	\end{figure}
	In Fig. \ref{lGraybody}, we plot the graybody factor $|T|^2$ for varing $l$. The decrease in the graybody factor with increasing $l$ correlates with the rise of peak value of the effective potential.
	
	\section{Test by Prony Method}\label{PronyTest}
		Finally, we used the Prony method \cite{Mining_Berti_2007} to further confirm the connection between our numerical calculations and the WKB approximation.
		In this section, the parameters are chosen as:
		$$\left\{ M \rightarrow \frac{1}{2} , a \rightarrow \frac{1}{5}, q \rightarrow \frac{1}{5}, \lambda \rightarrow \frac{2}{1000}\right\}\{l \rightarrow 2\}.$$
		\begin{figure}[!htb]
			\includegraphics [width=0.45\textwidth]{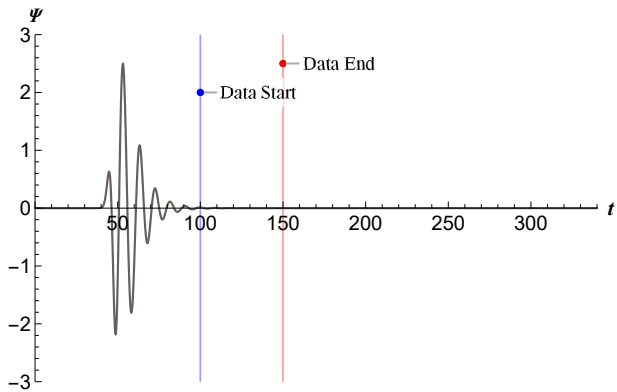}
			\caption{Time domain solution of  $\Psi$ over time at $V_{max}$.}
			\label{Data_a2q2lam20L2}
		\end{figure}
	
		We extract the data at $V_{max}$ in the interval $100<t<150$, as shown in Fig. \ref{Data_a2q2lam20L2}, and obtain the real and imaginary parts of the corresponding QNMs through the Prony method: $\omega_{Prony}=0.698097-0.121650i$. Correspondingly, the result obtained through the WKB approximation is $\omega_{WKB}=0.6980498152-0.1216230302i$.
		
\begin{figure}[!htb]
			\includegraphics [width=0.45\textwidth]{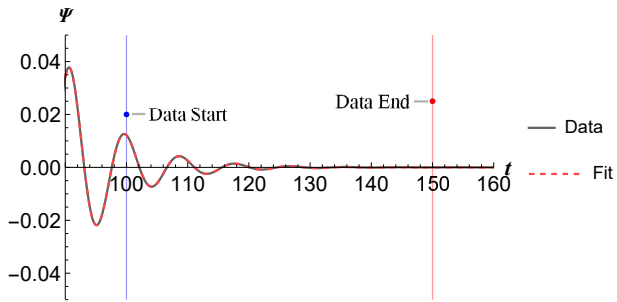}
			\caption{Fitted graph of the evolution of $\Psi$.}
			\label{Fit_a2q2lam20L2}
		\end{figure}

		Combined with Eq. \eqref{Pronyfit}, the fitting function can be written as:
		\begin{eqnarray}\label{Fitfun}
		  \Psi_{Prony} =  C_a \exp(-0.121650 t) \sin(0.698097  t + C_o)
		\end{eqnarray}
		where $ C_a = 2344.25 $ and $ C_o = -2.16557$. In Fig. \ref{Fit_a2q2lam20L2}, the red dashed line is the fitted curve Eq. \eqref{Fitfun}, and the black solid line is the data obtained by the numerical method.
		
		Finally, we can use the following error estimate method:
		\begin{eqnarray}
			\delta=\frac{\left|\omega_{WKB}-\omega_{Prony}\right|}{\left|\omega_{WKB}\right|}.
		\end{eqnarray}
		We obtain the corresponding error $\delta=7.67\times 10^{-5}$.
		This result in turn ensures the accuracy of our numerical calculations.	

	\section{Conclusion}\label{conclusion}
	In this paper, we investigated scalar perturbations in SBBH using the FEM and WKB approximation, unlike in \cite{Quasinormal_Fernando_2012}, We directly use the integral method to deal with the problem that this turtle coordinates have no analytic solution in the Bardeen space-time.
	Then, by numerical method, we obtain the slices of time evolution corresponding to QNMs of the scalar perturbation.
	We also calculated the corresponding QNMs and greybody factors using the WKB approximation. The results of  WKB scheme, the time domain diagram of the perturbation evolution and the effective potential are all related to each other. Finally, we further verified the accuracy of our calculations through the Prony method.
	
	The conclusions can be summarized as follows:
	
	1. In the time domain, the evolution of a scalar field does not increase with time. Correspondingly, in the WKB approximation, the imaginary part of the QNM frequency is always negative. This implies that the space-time under scalar perturbations is stable.
	
	2. As the parameter $a$ increases, the QNM frequency decreases while the decay rate slows down. This is because a larger parameter $a$ makes the effective potential smoother with lower peaks, due to the presence of string clouds weakening the effect of the original Bardeen background spacetime, making it flatter.
	
	3. Simultaneously, with an increase in parameter $a$, the greybody factor also increases proportionally. The presence of string clouds weakens the relative change of the effective potential, resulting in lower peak values and facilitating scalar field penetration through the potential barrier, thereby increasing the greybody factor.
	
	4. Conversely, an increase in the parameter $q$ leads to an increase in QNMs frequency, a slight reduction in decay rate, and a smaller greybody factor. It has almost no effect on the tail, because $q$ only affects the effective potential near the event horizon and can be ignored near the cosmological horizon.
	
	5. Increasing the cosmological constant $\lambda$ from zero results in a slight decrease in both the QNM frequency and decay rate. However, there is a significant modification in the tail behavior that is consistent with the notable impact of the cosmological constant $\lambda$ at large $r_*$ values.
	
	6. With an increase in angular quantum number $l$, the frequency of QNM increases, while the decay rate slightly decreases. Additionally, the grey body factor decreases correspondingly.
	
	It is worth noting that when $l=0$, a de Sitter phase occurs at $t\approx 320$ \cite{Conformal_Konoplya_2021}. We posit that this occurrence relates to the negative potential well that appears at $l=0$. At this point, bound states may exist, which produce a residual scalar field $\Psi_0$ in the tail. A simple analysis is available in Appendix \ref{Append2}.
	
	
	This work only studied scalar perturbations in SBBH spacetime. It is also interesting and straightforward to extend this work to vector and gravitational perturbations, especially for QNMs under gravitational perturbations, which reflect the fundamental characteristics of gravitational waves during the ringdown stage. These topics merit further exploration in the future.

	\begin{acknowledgments}
		This work is supported by National Natural Science Foundation of China with No. 12275087 and ``the Fundamental Research Funds for the Central Universities''.
	\end{acknowledgments}
	
	\appendix
	\section{Numerical Computation of Turtle Coordinates}\label{Append1}
	Firstly, before the numerical computation of turtle coordinates, we can divide the background spacetime into two categories and discuss them separately.
	
	The first category is when the cosmological constant $\lambda=0$. In this case, only the event horizon $r_h$ exists in the background spacetime, and there is no cosmological horizon. This means that when $r\rightarrow r_h$, $r_{*}\rightarrow -\infty$ and when $r\rightarrow +\infty$, $r_{*}\rightarrow +\infty$.
	
	The second category is when the cosmological constant $\lambda \neq 0$. In this case, both the event horizon $r_h$ and the cosmological horizon $r_c$ exist in the background spacetime. This means that when $r\rightarrow r_h$, $r_{*}\rightarrow -\infty$ and when $r\rightarrow r_c$, $r_{*}\rightarrow +\infty$.
	
	From the definition of turtle coordinates, we can obtain the integral function of $r_{*}(r)$:
	\begin{eqnarray}\label{intrstar}
		r_{*}(r) = \int_{r_0}^{r} \frac{1}{f(\bar{r})} d\bar{r},
	\end{eqnarray}
	where $r_0$ is an artificial selected integration parameter. In our calculation, when $\lambda=0$, we choose $r_0=2 r_h$. When $\lambda \neq 0$, we choose $r_0=(r_h + r_c)/2$. Note that the different choices of $r_0$ only cause a change in the integration constant $C_0$ and do not change the definition of the turtle coordinates.
	
	In the  integration, we encounter two problems. Firstly, the integral Eq. \eqref{intrstar} is a singular integral, and when $r_{*}(r)$ is close to the event horizon and cosmological horizon, it can usually be approximated by a logarithmic relationship:
	
	$r_* = C_h Log(r-r_h)+C_{h0}$ 	when $r\rightarrow r_h$ ;
		
	$r_* = - C_c Log(r_c-r)+C_{c0}$ when	$r\rightarrow r_c$.
		
	For example, if we need to compute the integral to $r_*(r)=-800$, let $r=r_h+\epsilon$, then we can get:
	\begin{eqnarray}
		-800 &=& C_h \ln(r_h+\epsilon-C_{h0})+C_h \notag \\
		&=& C_1 \ln(\epsilon)+C_{h0}\Leftrightarrow \epsilon=e^{{-(800+C_{h0})}/{C_h}}.
	\end{eqnarray}
	When the spacetime background parameters are selected as $a=1/5,M_1=0,M=1/2,q=2/5,\lambda=1/50$, we obtain the fitting parameters $C_h=2.202$ and $C_{h0}=-3.710$. At this point, the working accuracy requirement reaches $e^{{-(800-3.710)}/{2.202}}=e^{-361.621} \sim  10^{-158}$. It can be seen that the numerical integration precision requirement for Eq. \eqref{intrstar} is very high. Therefore, we need to pay attention to ensuring the working precision is large enough. At the same time, if the working precision is too high, it will lead to a sharp increase in computing resources. Therefore, we must choose an appropriate working precision for numerical integration according to the calculation demand.
	
	In addition, we only use the relationship between $r_*$ and $r$ in the time evolution equation. Furthermore, we usually require $r_*$ to be selected as a series of equidistant points. Obtaining the corresponding $r$ values for this series of equidistant turtle coordinates is not easy. Generally, we know the analytical relationship between $r_*$ and $r$, and then obtain the $r$ values  corresponding to a series of equidistant turtle coordinates.
	
	In this paper, because the analytical solution is unknown, and the working accuracy requirement is very high, the computational cost is substantial. Therefore, we adopt an interpolation method to obtain the $r$ values  corresponding to a series of equidistant turtle coordinate points. The steps are as follows:
	
	1. Select a series of appropriate $r$ values and integrate to obtain the corresponding $r_*$ values. For example, when the cosmological constant is non-zero, we divide the $r$ series values into two segments: $(r_h+\epsilon_h,r_0)$ and $(r_0,r_c-\epsilon_c)$. In the segment $(r_h+\epsilon_h,r_0)$, we choose the $r$ series values as $r_h+\epsilon_h+(r_0-r_h-\epsilon_h)/\epsilon_h^{(m-k)/m}$. In the segment $(r_0,r_c-\epsilon_c)$, we choose the $r$ series values as $r_c-\epsilon_c+(r_c-r_0-\epsilon_c)/\epsilon_c^{(m-k-1)/m}$. Here, $n$ is the number of points and $k=1,2,...,m$ is the series value.
	
	2. Perform integration for each $r$ value to obtain the corresponding $r_*$.
	
	3. Select a series of equidistant $r_*$ values within the range of the integrated $r_*$ values and use interpolation to obtain the corresponding $r$ values.

	When $m$ is sufficiently large, the error caused by interpolation becomes small enough to be negligible.
	Here, for different background spacetime parameters, we have chosen appropriate parameters $+\epsilon_h$ and $\epsilon_c$ such that $r_h+\epsilon_h$ and $r_c-\epsilon_c$ correspond to $r_{*min}\sim -800$ and $r_{*max}\sim 800$ and we set $m=40000$.
	
	\section{The wells of effective potential and de Sitter tails.}\label{Append2}
	We assume that after a substantial period of time, the evolution of the scalar field reaches a steady state, where $\Psi$ is fixed at constant values. This implies that $\Psi$ no longer varies with time, i.e., $\Psi_{j}^{i+1}=\Psi_{j}^{i}=\Psi_{j}$. Under this assumption, the difference Eq. \eqref{FDMFun} simplifies to:
	\begin{eqnarray} \label{FDMFunA}
		\Psi_{j}=
		&&-\Psi_{j}
		+\left(2-2 \frac{\Delta t^{2}}{\Delta r_{*}^{2}}-\Delta t^{2} V_{j}\right) \Psi_{j}   \notag \\
		&&+\frac{\Delta t^{2}}{\Delta r_{*}^{2}}\left(\Psi_{j-1}+\Psi_{j+1}\right) .
	\end{eqnarray}
	Further simplification yields the expression:
	\begin{eqnarray}\label{FDMFunB}
		\Psi_{j}  = \frac{1}{\left(2 +\Delta r_{*}^{2} V_{j}\right)}\left(\Psi_{j-1}+\Psi_{j+1}\right).
	\end{eqnarray}
	When $V_{eff}$ is always greater than zero, $\Psi_{j} <\left(\Psi_{j-1}+\Psi_{j+1}\right)/2$ forms a concave function. As $j\rightarrow \infty$, $\Psi_{j}\rightarrow \infty$, leading to instability. Therefore, if there exists a constant non-zero residual $\Psi_0$\cite{Radiative_Brady_1999,Late_Ismail_2021}, the effective potential must have negative values (i.e., a potential well) present.
	
	\bibliographystyle{unsrt}

\end{document}